\documentclass[preprint,showkeys,unsortedaddress,superscriptaddress,amsfonts,amsmath,amssymb]{revtex4}
\usepackage{graphicx}
\begin{document}
\title{Polytropic Inspired Inflation on the Brane}

\author{M. Setare}
\affiliation{Department of Science of Bijar, University of Kurdistan, Bijar, Iran}
\email{rezakord@ipm.ir}
\author{A. Ravanpak}
\affiliation{Department of Physics,
Vali-e-Asr University, Rafsanjan, Iran}
\email{a.ravanpak@vru.ac.ir}
\author{H. Farajollahi}
\affiliation{Department of Physics,
University of Guilan, Rasht, Iran}
\affiliation{School of Physics, University of New South Wales,
Sydney, NSW, 2052, Australia}
\email{hosseinf@guilan.ac.ir}

\date{\small {\today}}

\begin{abstract}

In this paper, the brane inflationary model inspired by polytropic inflationary idea is studied. In slow-roll approximation and high energy limit, for a chaotic potential, the model is developed and its characteristics is discussed. We obtain explicit expressions for the scalar power spectrum, the tensor-scalar ratio, the scalar spectral index and its running in terms of the polytropic parameters. We find a new constraint on the energy scale of the inflation and the brane tension using the WMAP9 data.

\end{abstract}


\maketitle

\section{Introduction}\label{sec1}

As is well known, many long-standing problems of the standard hot big bang scenario, such as flatness and horizon problems, can be removed if one consider a very rapidly accelerated expansion phase at the very early stages of the history of the universe which is called inflation \cite{1,2}.

The past few years witnessed a great deal of attention in higher dimensional gravity. It was shown that the fundamental higher-dimensional gravitational scale can be reduced down to the
TeV-scale in the presence of a large extra dimension \cite{3,nima,nima2}. In these theories the standard model of particles is confined to a lower-dimensional brane and only gravity can propagate into the bulk. The effect of the extra dimension induces additional terms in the Friedmann equation on the brane \cite{4,5,6}. Specially, the existence of a quadratic term in the energy density generally makes it easier to obtain inflation in the early universe \cite{7,8,9,10,11}.

On the other hand, the polytropic gas has recently been proposed as an alternative dark energy model for describing the late time acceleration of the universe \cite{12}, after that its equation of state is frequently used in various astrophysical situations, such as degenerate white dwarfs, neutron stars, main sequence stars \cite{13}, and also in the case of Lane-Emden models \cite{14,15}. Then, a question was raised; can we have a polytropic inspired inflation like a Chaplygin inspired inflation? Bertolami et al first proposed a phenomenological modification of gravity inspired by the Chaplygin gas equation of state which was then known as Chaplygin inspired inflation \cite{16}. In this scenario the standard inflaton field can be extrapolate to obtain a successful inflationary era with a Chaplygin gas model. After that the authors in \cite{17}, extended this procedure to a brane-world Chaplygin inspired
inflationary model. Similar works have been done in the context of a Chaplygin inspired inflation using a tachyon scalar field \cite{campo2,herrera2}. In a recent work a similar approach was used to introduce a polytropic inspired inflationary model \cite{setare}. This scenario was extended to the brane cosmology in which a tachyon field was considered as the inflaton \cite{setare2}. On the other hand in \cite{chavanis}, a generalized polytropic gas equation of state has been introduced and its cosmological applications has been investigated. In the present paper, assuming a canonical scalar field as the inflaton field we study the polytropic inflationary scenario in brane-world models. It can provide the conditions to compare with the results in \cite{setare} and \cite{setare2}. The outline of the paper is as follows. The next section presents a short review of the modified
Friedmann equation in brane-world model in the presence of a polytropic gas.
In Sec.\ref{sec3} we study the scalar and tensor perturbations in our
model and derive the cosmological parameters. In Sec.\ref{sec4} we use a
chaotic potential in the high-energy limit to obtain explicit
expressions for model parameters. Finally, in Sec.\ref{sec5} we present a
conclusion.

\section{The Model}\label{sec2}

In this section, a five-dimensional brane cosmology is studied with
the modified Friedmann equation given by
\begin{equation}\label{friedmann}
H^2=\kappa\rho_\phi[1+\frac{\rho_\phi}{2\lambda}]+\frac{\Lambda_4}{3}+\frac{\xi}{a^4},
\end{equation}
where $H = \dot{a}/a$ is the Hubble parameter. The $\rho_\phi$ and
$\Lambda_4$ are respectively, the matter field confined to the brane
and the four-dimensional cosmological constant. We also assume that
$\kappa = 8\pi G/3 = 8\pi/(3m_p^2)$. The influence of the bulk
gravitons on the brane is shown in the last term of the equation,
where $\xi$ is an integration constant. The four and five-dimensional
Planck masses are related through brane tension $\lambda$ in $m_p =
\sqrt{3M_5^6/(4\pi\lambda)}$, where constrained by nucleosynthesis.
The brane tension satisfies the inequality $\lambda
> (1MeV)^4$. We suppose that the four-dimensional cosmological constant is
zero. In addition, at the beginning of inflation, the last term
in (\ref{friedmann}) vanishes. Thus, the equation (\ref{friedmann})
reduces to
\begin{equation}\label{simplefriedmann}
H^2=\kappa\rho_\phi[1+\frac{\rho_\phi}{2\lambda}].
\end{equation}

On the other hand, the polytropic fluid which has been proposed as an
alternative model for describing the accelerating of the universe \cite{12},
has an equation of state of the form
\begin{equation}\label{eos}
p=K\rho^{1+\frac{1}{n}},
\end{equation}
where $K$ and $n$ are constant values called in the literature
polytropic constant and polytropic index, respectively. The polytropic
constant $K$ can take the positive value for radiation and stiff fluid, the zero value for
dust and the negative value for inflationary scenario \cite{12}. The case $n=-\frac{1}{2}$, is motivated by string theory \cite{Randall,Dvali,Freese,Maartens,Lue}.
Now, using (\ref{eos}) and conservation equation
\begin{equation}\label{conservation}
\dot{\rho}+3H(\rho+p)=0,
\end{equation}
we can obtain
\begin{equation}\label{polytropic}
\rho=[-K+Ba^{\frac{3}{n}}]^{-n},
\end{equation}
where $B$ is an integration constant. In (\ref{conservation}) dot means derivative with respect to the cosmological time.

In this article we will not consider
the above equation as a consequence of the polytropic equation of state (\ref{eos}), but we start by studying
the modified gravity, where the gravitational dynamic is given by modified Friedmann equation as \cite{Marr}
\begin{equation}\label{generalfried}
H^2=\kappa[-K+{\rho_\phi}^{\frac{-1}{n}}]^{-n}[1+\frac{([-K+{\rho_\phi}^{\frac{-1}{n}}]^{-n})}{2\lambda}].
\end{equation}
In here, $\rho_\phi=\frac{1}{2}\dot{\phi}^2+V(\phi)$ and $V(\phi)$ is the scalar potential. This modification of energy density
is understood from an extrapolation of equation (\ref{polytropic}) as
\begin{equation}\label{extrapolate}
    \rho=[-K+{\rho_m}^{\frac{-1}{n}}]^{-n}\rightarrow [-K+{\rho_\phi}^{\frac{-1}{n}}]^{-n}
\end{equation}
where $\rho_m$ is the matter energy density \cite{setare,setare2,setare3}.

The Klein-Gordon equation for the scalar field confined on the brane
is written as
\begin{equation}\label{KG}
\ddot{\phi}+3H\dot{\phi}+V'=0.
\end{equation}
and the pressure of a scalar field is expressed as
\begin{equation}\label{pressure}
p_\phi=\frac{1}{2}\dot{\phi}^2-V(\phi).
\end{equation}
During the inflationary epoch the energy density associated to the
scalar field is of the order of the potential, i.e. $\rho_\phi\sim
V$. Assuming the set of slow-roll conditions, i.e. $\dot{\phi}^2<<
V(\phi)$ and $\ddot{\phi}<<V'$, the Friedmann equation
(\ref{generalfried}) reduces to
\begin{equation}\label{fried}
H^2\simeq
\kappa[-K+V^{\frac{-1}{n}}]^{-n}[1+\frac{([-K+V^{\frac{-1}{n}}]^{-n})}{2\lambda}].
\end{equation}
Also equation (\ref{KG}) becomes
\begin{equation}\label{simpleKG}
3H\dot{\phi}\simeq -V'.
\end{equation}
Introducing the dimensionless slow-roll parameters, we can write
\begin{equation}\label{slowroll1}
\varepsilon = -\frac{\dot H}{H^2} \simeq
\frac{V'^2}{6\kappa}V^{\frac{-1}{n}-1}[-K+V^{\frac{-1}{n}}]^{n-1}\frac{[1+\frac{([-K+V^{\frac{-1}{n}}]^{-n})}
{\lambda}]}{[1+\frac{([-K+V^{\frac{-1}{n}}]^{-n})}{2\lambda}]^2},
\end{equation}
and $\delta=\frac{-\ddot{\phi}}{H\dot{\phi}}=\frac{\dot{H}}{H^2}+\frac{V''}{3H^2}$
as
\begin{equation}\label{slowroll2}
\delta \simeq
\frac{-V'^2}{6\kappa}V^{\frac{-1}{n}-1}[-K+V^{\frac{-1}{n}}]^{n-1}\frac{[1+\frac{([-K+V^{\frac{-1}{n}}]^{-n})}
{\lambda}]}{[1+\frac{([-K+V^{\frac{-1}{n}}]^{-n})}{2\lambda}]^2}+\frac{V''}{3\kappa[-K+V^{\frac{-1}{n}}]^{-n}
[1+\frac{([-K+V^{\frac{-1}{n}}]^{-n})}{2\lambda}]}.
\end{equation}
Also, we can define
\begin{equation}\label{slowroll3}
\eta\equiv\varepsilon+\delta=\frac{V''}{3\kappa[-K+V^{\frac{-1}{n}}]^{-n}
[1+\frac{([-K+V^{\frac{-1}{n}}]^{-n})}{2\lambda}]},
\end{equation}
where in the low-energy limit $(-K+\rho_\phi^{-1/n})^{-n}\ll\lambda$, they reduce to the standard form in \cite{setare}.

The condition under which inflation takes place can be summarized
with the parameter $\varepsilon$ satisfying the inequality
$\varepsilon<1$, which is analogue to the requirement that
$\ddot{a}>0$. This condition could be written in terms of the
scalar potential and its derivative $V$ and $V'$, as
\begin{equation}\label{conditionofinflation}
V'^2V^{\frac{-1}{n}-1}[1+\frac{([-K+V^{\frac{-1}{n}}]^{-n})}
{\lambda}]<6\kappa([-K+V^{\frac{-1}{n}}]^{1-n})[1+\frac{([-K+V^{\frac{-1}{n}}]^{-n})}{2\lambda}]^2.
\end{equation}
Inflation ends when the universe heats up at a time when
$\varepsilon\simeq1$, which implies
\begin{equation}\label{endofinflation}
V_f'^2V_f^{\frac{-1}{n}-1}[1+\frac{([-K+V_f^{\frac{-1}{n}}]^{-n})}
{\lambda}]\simeq
6\kappa([-K+V_f^{\frac{-1}{n}}]^{1-n})[1+\frac{([-K+V_f^{\frac{-1}{n}}]^{-n})}{2\lambda}]^2,
\end{equation}
where the subscript $f$, has been used for the final values. In high energies when
$[-K+V^{\frac{-1}{n}}]^{-n}\gg\lambda$, equation
(\ref{endofinflation}) becomes
\begin{equation}\label{highend}
V_f'^2\simeq
\frac{3\kappa}{2\lambda}V_f^{\frac{1}{n}+1}[-K+V_f^{\frac{-1}{n}}]^{1-2n}.
\end{equation}
Also, the number of e-folds at the end of inflation where indicates the count that scale factor parameter grows by a factor $e$, is given by
\begin{equation}\label{efold}
N = \int Hdt = \int\frac{H}{\dot\phi}d\phi =
-3\int_{\phi_\ast}^{\phi_f}\frac{H^2}{V'}d\phi =
-3\kappa\int_{\phi_\ast}^{\phi_f}\frac{[-K+V^{\frac{-1}{n}}]^{-n}}{V'}[1+\frac{(-K+V^{\frac{-1}{n}})^{-n}}{2\lambda}]d\phi,
\end{equation}
in which $\ast$, stands for the epoch when the cosmological scales exit the horizon. Also, we can rewrite Eq.(\ref{efold}) in the high-energy limit as
\begin{equation}\label{highefold}
N = -\frac{3\kappa}{2\lambda}\int_{V_\ast}^{V_f}\frac{[-K+V^{\frac{-1}{n}}]^{-2n}}{{V'}^2}dV.
\end{equation}

\section{Perturbations}\label{sec3}

In this section we will study the scalar and tensor perturbations
for our model following the procedure used in \cite{18}--\cite{20}. It was shown in \cite{Wands} that the curvature perturbation $\cal{R}$ on comoving hypersurfaces is related to the field fluctuation $\delta\phi$ via
\begin{equation}\label{ops}
    {\cal{R}}=\frac{H}{\dot\phi}\delta\phi=\frac{H}{\dot\phi}(\frac{H}{2\pi}).
\end{equation}
Thus, for a scalar field the power spectrum of the
curvature perturbations is given in the slow-roll approximation by the expression ${\cal{P}_R} = (\frac{H^2}{2\pi\dot\phi})^2$ \cite{7}. So, in our model
\begin{equation}\label{scalarperturb}
{\cal{P}_R} \simeq
\frac{9\kappa^3}{4\pi^2V'^2}[-K+V^{\frac{-1}{n}}]^{-3n}[1+\frac{([-K+V^{\frac{-1}{n}}]^{-n})}{2\lambda}]^3.
\end{equation}
Since according to standard notation in the literature the power spectrum is the contribution to the variance per unit
logarithmic interval in the wave number $k$, to describe the slope of the power-spectrum we define the scalar spectral index $n_s$, through $n_s-1 =
\frac{d\ln{\cal{P}_R}}{d\ln k}$, where the interval in wave number
is related to the number of e-folds by the relation $d\ln k(\phi) =
dN(\phi)$. From Eq.(\ref{scalarperturb}), we get, $n_s =
1-4\varepsilon+2\delta = 1-6\varepsilon+2\eta$, or equivalently
\begin{equation}\label{ns}
n_s =
1-\frac{(-K+V^{\frac{-1}{n}})^n}{3\kappa}[1+\frac{(-K+V^{\frac{-1}{n}})^{-n}}{2\lambda}]^{-1}(\frac{3{V'}^2V^{\frac{-1}{n}-1}}{(-K+V^{\frac{-1}{n}})}
\frac{[1+\frac{(-K+V^{\frac{-1}{n}})^{-n}}{\lambda}]}{[1+\frac{(-K+V^{\frac{-1}{n}})^{-n}}{2\lambda}]}-2V'').
\end{equation}

Planck 2015 results \cite{Ade}, indicate that there is not any statistically significant running of the scalar spectral index $n_{run} =
dn_s/d\ln k$, but we can obtain its relation in our model as
\begin{equation}\label{nrun}
n_{run} =
\frac{4(-K+V^{\frac{-1}{n}})}{V^{\frac{-1}{n}-1}V'}\frac{[1+\frac{(-K+V^{\frac{-1}{n}})^{-n}}{2\lambda}]}{[1+\frac{(-K+V^{\frac{-1}{n}})^{-n}}{\lambda}]}
[3\varepsilon'-\eta']\varepsilon.
\end{equation}

On the other hand, the generation of tensor perturbations during
inflation would produce gravitational waves \cite{Bassett} and these perturbations
in brane cosmology are more involved since gravitons can propagate into the
bulk. The amplitude of tensor perturbations in brane scenario has been calculated in literature \cite{Langlois, Sami}
\begin{equation}\label{tensorperturb}
{\cal P}_g=24\kappa(\frac{H}{2\pi})^2F^2(x),
\end{equation}
where $x = Hm_p\sqrt{3/(4\pi\lambda)}$ and
\begin{equation}\label{fx}
F(x)=[\sqrt{1+x^2}-x^2\sinh^{-1}(\frac{1}{x})]^{-\frac{1}{2}}.
\end{equation}
In our model it reduces to
\begin{equation}\label{tensorperturb2}
    {\cal P}_g\simeq
\frac{6\kappa^2}{\pi^2}[-K+V^{\frac{-1}{n}}]^{-n}[1+\frac{([-K+V^{\frac{-1}{n}}]^{-n})}{2\lambda}]F^2(x),
\end{equation}
From expressions (\ref{scalarperturb}) and (\ref{tensorperturb2}) we
write the tensor-scalar ratio as
\begin{equation}\label{ratio}
r=(\frac{{\cal P}_g}{{\cal P}_{\cal R}})|_{k=k_\ast} \simeq
\frac{8{V'}^2}{3\kappa}\frac{(-K+V^{\frac{-1}{n}})^{2n}}{(1+\frac{(-K+V^{\frac{-1}{n}})^{-n}}{2\lambda})^2}F^2(V)|_{k=k_\ast}
\cdot
\end{equation}
Here, $k_\ast$ is referred to $k = Ha$, the value when the universe
scale crosses the Hubble horizon during inflation.

\section{Chaotic potential in the high energy limit}\label{sec4}

Let us consider an inflaton scalar field $\phi$, with a chaotic
potential as $V = m^2\phi^2/2$, where $m$ is the mass of the scalar field. In the following, we
develop our model in the high-energy limit, i.e.
$(-K+V^{\frac{-1}{n}})^{-n}\gg\lambda$.

Since for the potential under consideration, ${V'}^2 = 2m^2V$, using Mathematica we obtain from Eq.(\ref{highefold})
\begin{equation}\label{finalefold}
    N=\frac{3\kappa}{8m^2\lambda}[h(V_\ast)-h(V_f)],
\end{equation}
in which
\begin{equation}\label{h}
    h(V)=V^2 {_2F_1}(2n,2n;2n+1;KV^{\frac{1}{n}}),
\end{equation}
and $_2F_1$, is a hypergeometric function.
Also, Eq.(\ref{scalarperturb}) reduces to
\begin{equation}\label{highscalarperturb}
{\cal{P}_R} \approx
\frac{9\kappa^3}{64\pi^2m^2\lambda^3V}(-K+V^{\frac{-1}{n}})^{-6n},
\end{equation}
and from Eq.(\ref{ns}) and Eq.(\ref{nrun}) we obtain
\begin{equation}\label{highns}
n_s \approx
1-\frac{4m^2\lambda}{3\kappa}(-K+V^{\frac{-1}{n}})^{2n}(\frac{6V^{\frac{-1}{n}}}{(-K+V^{\frac{-1}{n}})}-1),
\end{equation}
and
\begin{equation}\label{highnrun}
n_{run} \approx
-\frac{32m^4\lambda^2}{9n\kappa^2}(-K+V^{\frac{-1}{n}})^{4n-2}[(n-3)K+5nV^{\frac{-1}{n}}]V^{\frac{-1}{n}},
\end{equation}
respectively.

In the case of tensor perturbations, one can check that in the high energy limit, we have $F^2(x)\approx\frac{3}{2}x$ \cite{lidsey,naciri}, where in our model $x=\frac{1}{\lambda}[-K+V^{\frac{-1}{n}}]^{-n}$. Then, the power spectrum of tensor perturbation can be rewritten as
\begin{equation}\label{finalpg}
    {\cal{P}}_g=\frac{9\kappa^2}{2\pi^2\lambda^2}[-K+V^{\frac{-1}{n}}]^{-3n},
\end{equation}
and then the tensor to scalar ratio becomes
\begin{equation}\label{highratio}
r \approx
\frac{32m^2\lambda V}{\kappa}(-K+V^{\frac{-1}{n}})^{3n}.
\end{equation}

Using Eqs.(\ref{highscalarperturb}) and (\ref{highratio}), one can solve $K$ and $m$, in terms of $\cal{P}_R$, $r$, $V$ and $\lambda$ as follows
\begin{equation}\label{lambda}
K =V^{\frac{-1}{n}}-[\frac{2\pi^2\lambda^2{\cal{P}_R}r}{9\kappa^2}]^{\frac{-1}{3n}},
\end{equation}
and
\begin{equation}\label{mass}
m =\frac{\pi r}{12}\sqrt{\frac{\lambda {\cal{P}_R}}{\kappa V}}.
\end{equation}
As we mentioned in Sec.II, we know from \cite{12} that for an inflationary universe $K<0$. Thus, from (\ref{lambda}), the ratio $\frac{V^3}{\lambda^2}$, satisfies the inequality below
\begin{equation}\label{inequality}
    \frac{V^3}{\lambda^2}>\frac{2\pi^2r{\cal{P}_R}}{9\kappa^2},
\end{equation}
where according to the results of WMAP nine year data \cite{wmap}, if we choose ${\cal{P}_R}=2.464\times10^{-9}$ and $r=0.13$, a lower limit for this ratio can be obtained as $V_\ast^3/\lambda^2>7.02\times10^{-10}$, in which we have assumed $\kappa=1$. We should note that in \cite{17} a lower limit, $V_\ast^3/\lambda^2>7.2\times10^{-11}m_p^4$, and in \cite{herrera2} an upper limit, $V_\ast^3/\lambda^2<2.4\times10^{-11}m_p^4$, for this ratio has been obtained but using WMAP5 data and in a Chaplygin inspired brane inflationary scenario, using respectively a standard and a tachyon scalar field. For comparison, we can rewrite our result in terms of $m_p^4$, as $V_\ast^3/\lambda^2>1.001\times10^{-11}m_p^4$.

\section{Summary}\label{sec5}

In this paper we studied polytropic inflationary model in the context of a braneworld scenario. We obtained the dimensionless slow-roll parameters. The condition under which inflation takes place is discussed by using the parameter $\varepsilon$ satisfying the inequality $\varepsilon<1$. This condition is expressed in terms of the scalar potential and its derivative as given by Eq.(\ref{conditionofinflation}). Then, we studied the scalar and tensor perturbations and derived the power spectrum of the curvature perturbations in the slow-roll approximation and an expression for the corresponding scalar spectrum index $n_s$ and its running $n_{run}$. Then, we considered an inflaton scalar field with a chaotic potential and developed the model in the high-energy limit. We obtained the scalar power spectrum, the tensor-scalar ratio, the scalar spectral index and its running in terms of polytropic constant $K$, polytropic index $n$, mass of the inflaton field $m$, brane tension $\lambda$ and the potential $V$ . By using these relations and observational data from WMAP9 we obtained a new lower limit for the ratio $V^3/\lambda^2$.

\end{document}